\documentclass{aa}  

\usepackage{graphicx}
\usepackage{subfig}
\usepackage{enumitem}
\usepackage{txfonts}
\usepackage{hyperref}
\usepackage{mathtools}

\begin{document} 

\title{On the importance of special relativistic effects in modelling ultra-fast outflows}
\author{
A. Luminari\inst{1,2},
F. Tombesi\inst{1,3,4,2},
E. Piconcelli\inst{2},
F. Nicastro\inst{2},
K. Fukumura\inst{5},
D. Kazanas\inst{4},
F. Fiore\inst{6},
L. Zappacosta\inst{2}
}

\institute{
Department of Physics, University of Rome ``Tor Vergata'', Via della Ricerca Scientifica 1, I-00133 Rome, Italy
\and
INAF - Osservatorio Astronomico di Roma, Via Frascati 33, 00078 Monteporzio, Italy
\and
Department of Astronomy, University of Maryland, College Park, MD 20742, USA
\and
NASA/Goddard Space Flight Center, Code 662, Greenbelt, MD 20771, USA
\and
Department of Physics and Astronomy, James Madison University, Harrisonburg, VA 22807, US
\and
INAF - Osservatorio Astronomico di Trieste, via G.B. Tiepolo 11, 34131, Trieste, Italy
}
\date{Received 27/09/19; accepted 25/11/19}

\abstract
{Outflows are observed in a variety of astrophysical sources. Remarkably, ultra-fast ($v\geq 0.1c$), highly ionised outflows in the UV and X-ray bands are often seen in Active Galactic Nuclei (AGNs). Depending on their kinetic power and mass outflow rate, respectively $\dot{E}_{out}, \dot{M}_{out}$, such outflows may play a key role in regulating the AGN-host galaxy co-evolution process through cosmic time and metal-feeding the surrounding CGM/IGM. It is therefore crucial to provide accurate estimates of the wind properties, including $\dot{M}_{out}, \dot{E}_{out}$.}
{Here, we concentrate on special relativistic effects concerning the interaction of light with matter moving at relativistic speed relatively to the source of radiation. Our aim is to assess the impact of these effects on the observed properties of the outflows and implement a correction for these effects in the existing spectral modelling routines.}
{We define a simple procedure to incorporate relativistic effects in radiative transfer codes. Following this procedure, we run a series of simulations to explore the impact of relativistic effects for different outflow velocities and column densities.}
{The observed optical depth of the wind is usually considered a proxy for its column density $N_H$, independently on the velocity of the outflow. However, our simulations show that the observed optical depth of an outflow with a given column density $N_H$ decreases rapidly as the velocity of the wind approaches relativistic values. This, in turn, implies that when estimating $N_H$ from the optical depth, it is necessary to include a velocity-dependent correction, already for moderate velocities (e.g. $v_{out} \buildrel > \over \sim 0.05 c$). This correction linearly propagates to the derived quantities $\dot{M}_{out}, \dot{E}_{out}$.}
{We demonstrate that special relativistic effects must be considered in order to obtain correct estimates of $\dot{M}_{out}$ and $\dot{E}_{out}$ for an outflow moving at mildly relativistic speed relatively to the illuminating source of radiation. As an example we calculate the relativistically corrected values of $\dot{M}_{out}$ and $\dot{E}_{out}$ for a sample of $\sim 30$ Ultra-Fast Outflows (UFOs) taken from the literature and find correction factors of $20-120 \%$ within the observed range of outflowing velocities ($v_{out}\approx 0.1-0.3 c$). This brings the ratio between $\dot{M}_{out}$ and the disk accretion rate close or even above unity for the vast majority of the sources of the sample, highlighting the importance of the reported relativistic corrections to understand the growth of the most massive black holes. The upcoming next generation of high sensitivity X-ray telescopes such as {\it XRISM} and {\it Athena} will provide a much more complete census of UFOs, especially in the fastest velocity regime where the relativistic corrections are increasingly important.}


\titlerunning{Special relativistic effects in ultra-fast outflows}
\authorrunning{A. Luminari et al.}
\maketitle

\section{Introduction}
Outflows are ubiquitously observed from a variety of astrophysical sources and their impact on the surrounding environment depends on their energetic. In particular, mildly relativistic and ionised outflows from the innermost regions of Active Galactic Nuclei (AGNs) are often seen in UV and X-ray absorption spectra (e.g., \citealp{Chartas, T10, R11, B19}) and may carry sufficient energy to regulate both the growth of the central super-massive black hole (SMBH) and the evolution of the surrounding host galaxy (\citealp{C14, F12, T15, Z12}). This critically depends on the kinetic power of these outflows, which in turn depends on both their velocity and mass flux (\citealp{Dimatteo, KP15}). 

The line-of-sight velocity is typically inferred via the blue-shift of the absorption features imprinted by the outflowing material onto the continuum emission of the central source, compared to the systemic redshift of the host galaxy. The mass outflow rate $\Dot{M}_{out}$, instead, for a given covering factor and distance of the outflow, is estimated by measuring the optical depth of the absorption features. The observed optical depth is considered a proxy of the outflow column density $N_H$ along the line of sight, independently on its outflow velocity $v_{out}$.  

In this work we show that this assumption no longer holds for outflows escaping the central continuum source of radiation with velocities corresponding to a fraction of the speed of light $c$ (e.g. $v_{out} \buildrel > \over \sim 0.05 c$). For such outflows, the observed (i.e. apparent) optical depth of the spectral features produced by the absorbing material, significantly underestimates the intrinsic $N_H$ and, consequently, the mass transfer rate of the outflows. Therefore, a velocity-dependent correction must be adopted to account for this effect in the estimate of $N_H$.

This pure special-relativistic effect is universal (i.e. applies to any fast-moving line-of-sight outflow), and affects not only our estimate of the kinetic power of the outflow but also the ability of the radiative source to effectively accelerate the outflow outwards.
For AGN outflows, this may have deep implications on the feedback mechanism and the co-evolution with respect to the host galaxy (\citealp{KH13}).

The paper is organised as follows. In Sect. \ref{physics} we overview the special relativity treatment for a fast-moving gas embedded in a radiation field. In Section \ref{prescription} we show how to incorporate such treatment in modelling outflow spectra. In Section \ref{conclusions} we discuss the results and their implications on estimating $\dot{M}_{out}, \dot{E}_{out}$, and we summarise in Sect. \ref{sect5}.

\section{Special Relativistic Transformation in the Outflow Reference Frame}
\label{physics}
\begin{figure}
\includegraphics[width=9.5cm]{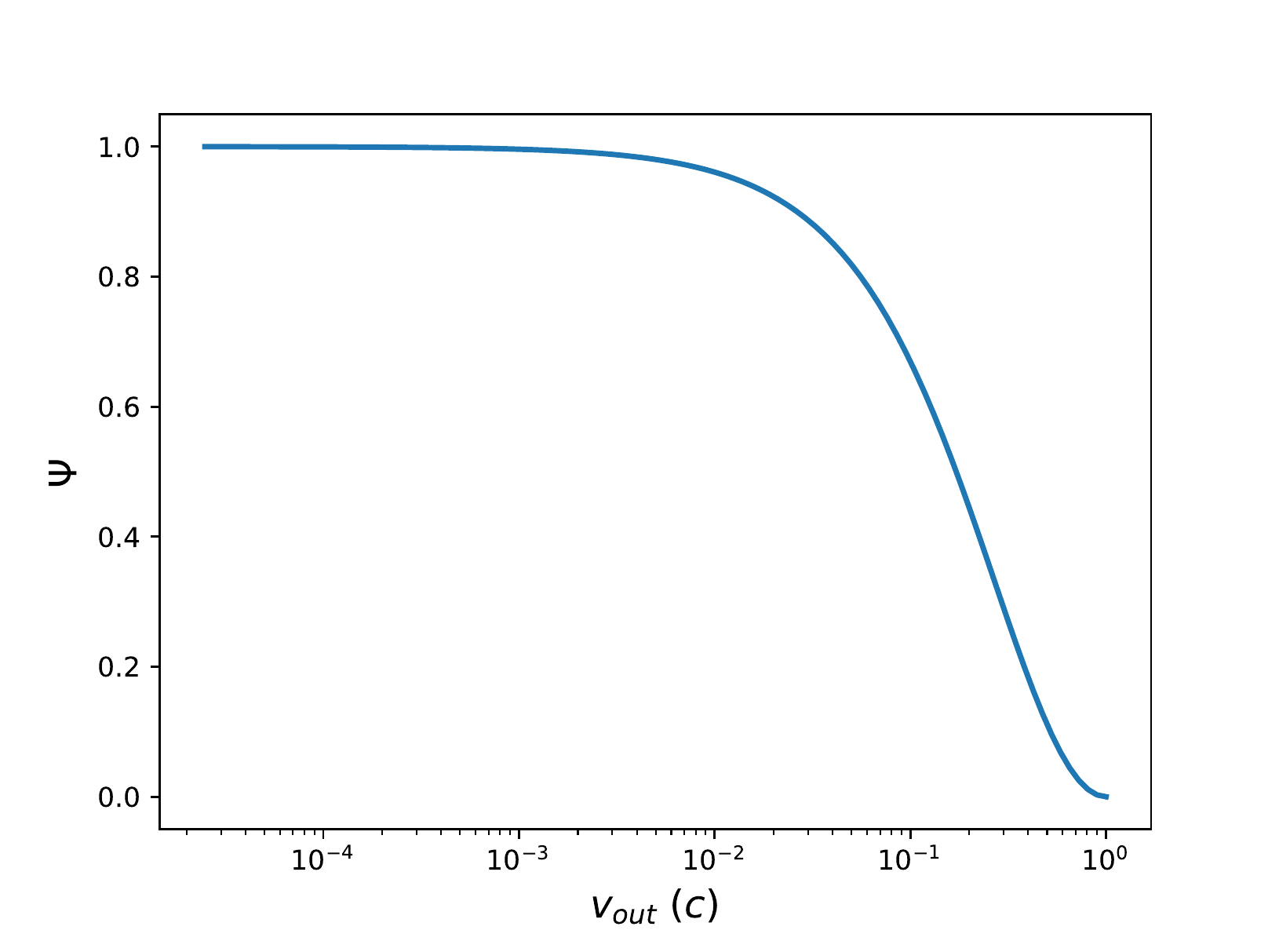}
\caption{Deboosting factor $\Psi$ in the gas reference frame $K'$ as a function of $v_{out}$ assuming $\theta=180$. For speeds lower than 0.1 the speed of light the radiation intercepted by the outflow and the (rest-frame) observer at infinity are virtually the same. For higher speeds, the fraction of intercepted radiation drops dramatically due to special relativistic effects. }
\label{psi}
\end{figure}

According to special relativity, the luminosity $L'$ seen by a clump of gas moving at relativistic speed is reduced of a factor $\Psi$, with respect to a static gas, as follows:
\begin{equation}
L'=L\cdot \Psi
\label{main}
\end{equation}
where $L$ is the luminosity seen by an observer at rest and $\Psi$, i.e. the de-boosting factor, is defined as:
\begin{equation}
\Psi\equiv\psi^4= \frac{1}{\gamma^4 (1-\beta cos(\theta))^4}
\label{main_long}
\end{equation}
where $\gamma \equiv \frac{1}{\sqrt{1-\beta^2}}$, $\beta=v_{out}/c$, $v_{out}$ is the gas velocity and $\theta$ is the angle between the incident luminosity $L$ and the direction of motion of the gas. Figure \ref{psi} shows $\Psi$ as a function of $v_{out}$ for $\theta=180\ deg$, corresponding to a radial outward motion of the gas. The deboosting factor is due to the combination of the space-time dilatation in the gas reference frame, $K'$, and the relativistic Doppler shift of the received radiation (\citealp{RL}).

Using Eq. \ref{main}, the radiative intensity (i.e., the luminosity per solid angle) $\frac{dL'}{d\Omega'}$ received by the outflowing gas in $K'$ can be written as a function of the intensity in the rest frame $K$, as follows:
\begin{equation}
\frac{dL'}{d\Omega'}= \Psi \frac{dL}{d\Omega} =\psi dE\cdot \psi^3\frac{1}{dt d\Omega}
\label{expl}
\end{equation}
where $dE, dt, d\Omega$ corresponds to the energy, time and solid angle intervals in $K$. Specifically, in Eq. \ref{expl}, $\psi dE$ is the energy transformation term, which represents the Doppler shift of the wavelengths in $K'$. The second term, $\psi^3\frac{1}{dt d\Omega}$, indicates a reduction of the intensity due to the space-time dilatation in $K'$.

Noteworthy, Eq. \ref{main} and \ref{expl} also describe the emission from gas moving at relativistic velocity, as usually observed in high velocity systems such as jets in Blazars and GRBs (\citealp{Urry, G93}). When radiation is emitted along the direction of motion, i.e. $\theta\approx0\ deg$, $\Psi$ increases with increasing $v_{out}$, while $\Psi\leq1$ when it is emitted perpendicularly or backward ($\theta=90\ deg$ and $180\ deg$, respectively). The overall result is to concentrate the emitted radiation into a narrow cone along the direction of motion, an effect known as "relativistic beaming" (\citealp{RL, EHT}).

Another way of describing the reduction of the luminosity seen by the outflowing gas is the following. In $K'$, the luminosity source appears as moving away with velocity $v_{out}$ and $\theta=180\ deg$ (for a pure radial motion), which results into a de-boosting of the received luminosity due to the relativistic beaming, according to Eq. \ref{main}.

\section{Modelling Outflow Absorption Spectra Including Special Relativistic Effects}
\label{prescription}
We propose to include these special relativistic corrections in modelling spectral absorption features from the outflowing gas, according to the following procedure (see Appendix \ref{appendix1} for a detailed description).

\begin{itemize}
\item{The first step is to transform the incident spectrum $S_I(K)$ from $K$ to $K'$, obtaining $S_I(K')$, according to Eq. \ref{expl}.}
\item{$S_I(K')$ is then given as input to the radiative transfer code to calculate the transmitted spectrum in the outflowing gas frame $K'$, $S_T(K')$.}
\item{Finally, the relativistic-corrected transmitted spectrum in $K$, i.e. $S_{out}(K)$, is given by:
\begin{equation}
S_{out}(K)=S_I(K)\cdot \Delta + S_T(K')\cdot \psi^{-1}
\label{sout}
\end{equation}
where $\Delta\equiv 1-\psi^3$. The term $S_T(K')\cdot \psi^{-1}$ indicates the spectrum  $S_T(K')$ in Doppler-shifted (from $K'$ to $K$) frequencies.}
\end{itemize}
We note that in the low-velocity limit $v_{out}\ll c$, $\Psi \approx 1, \Delta\approx0$ and the resulting spectrum is $S_{out}(K)=S_T(K')\cdot \psi^{-1}$, as it is usually calculated. For the opposite high-velocity regime $v_{out}\rightarrow c$, $\Psi\approx 0$ and the outflowing gas does not interact with the ionising radiation. In fact, $S_I(K')$ and $S_T(K')$ have null intensity (see Eq. \ref{expl}), $\Delta\approx 1$ and $S_{out}(K) \approx S_I(K)$.

We use the radiative transfer code \textit{XSTAR}, v2.5 (\citealp{xstar}) to calculate $S_{out}(K)$, which is the spectrum as seen by a rest frame observer in $K$. 

\begin{figure*}
\includegraphics[width=19cm]{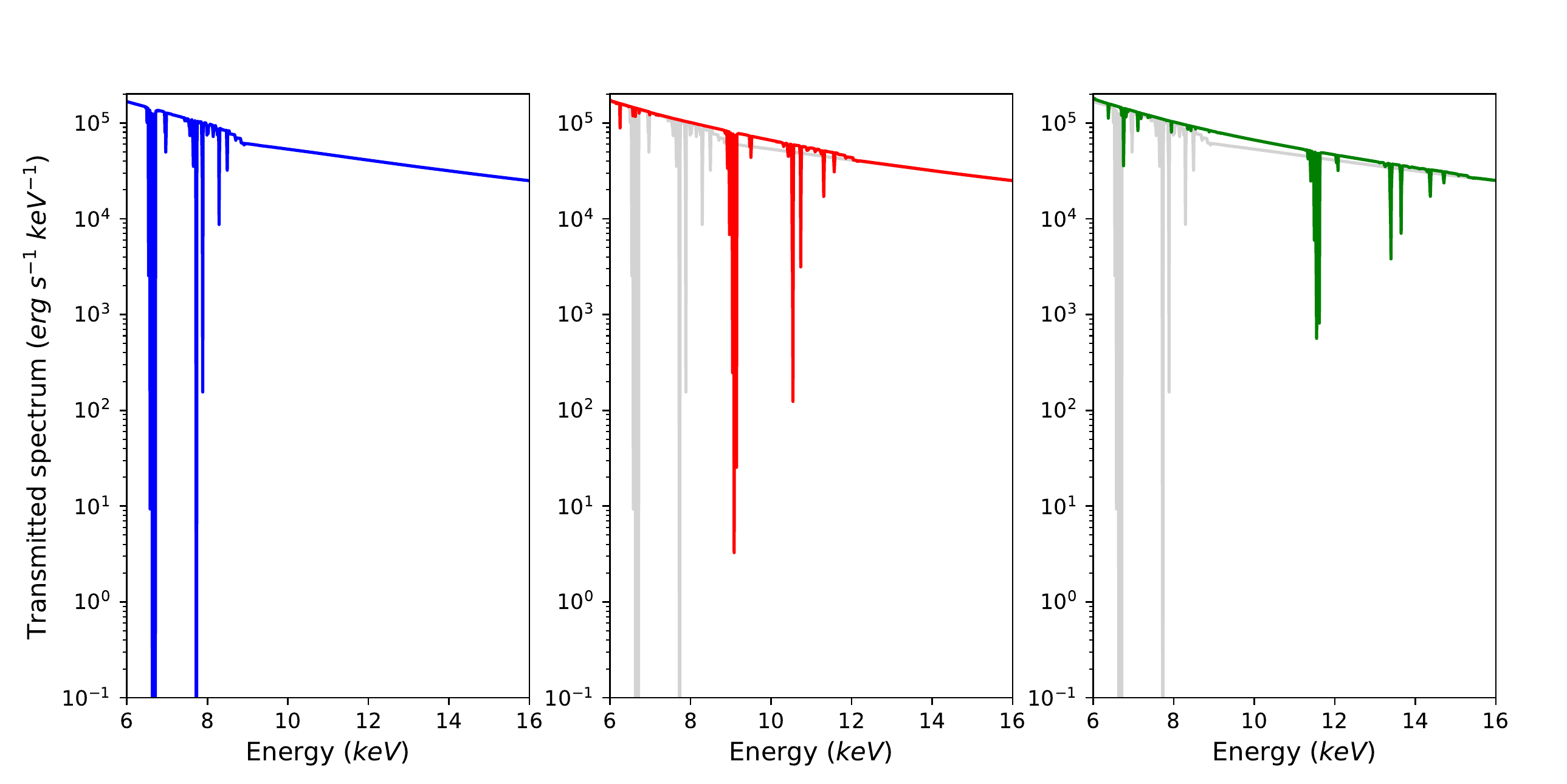}
\caption{Absorption spectra for increasing outflow velocity: $v_{out}=0.00$ (left panel), $=0.30$ (centre panel), $=0.50\ c$ (right panel). For comparison, in the centre and right panels we report (light grey) the absorption spectrum for $v_{out}=0.00\ c$. See Sect. \ref{prescription} for details on the spectral parameters used in this simulation.}
\label{abs_spectra}
\end{figure*}

\begin{figure}
\includegraphics[width=9.5cm]{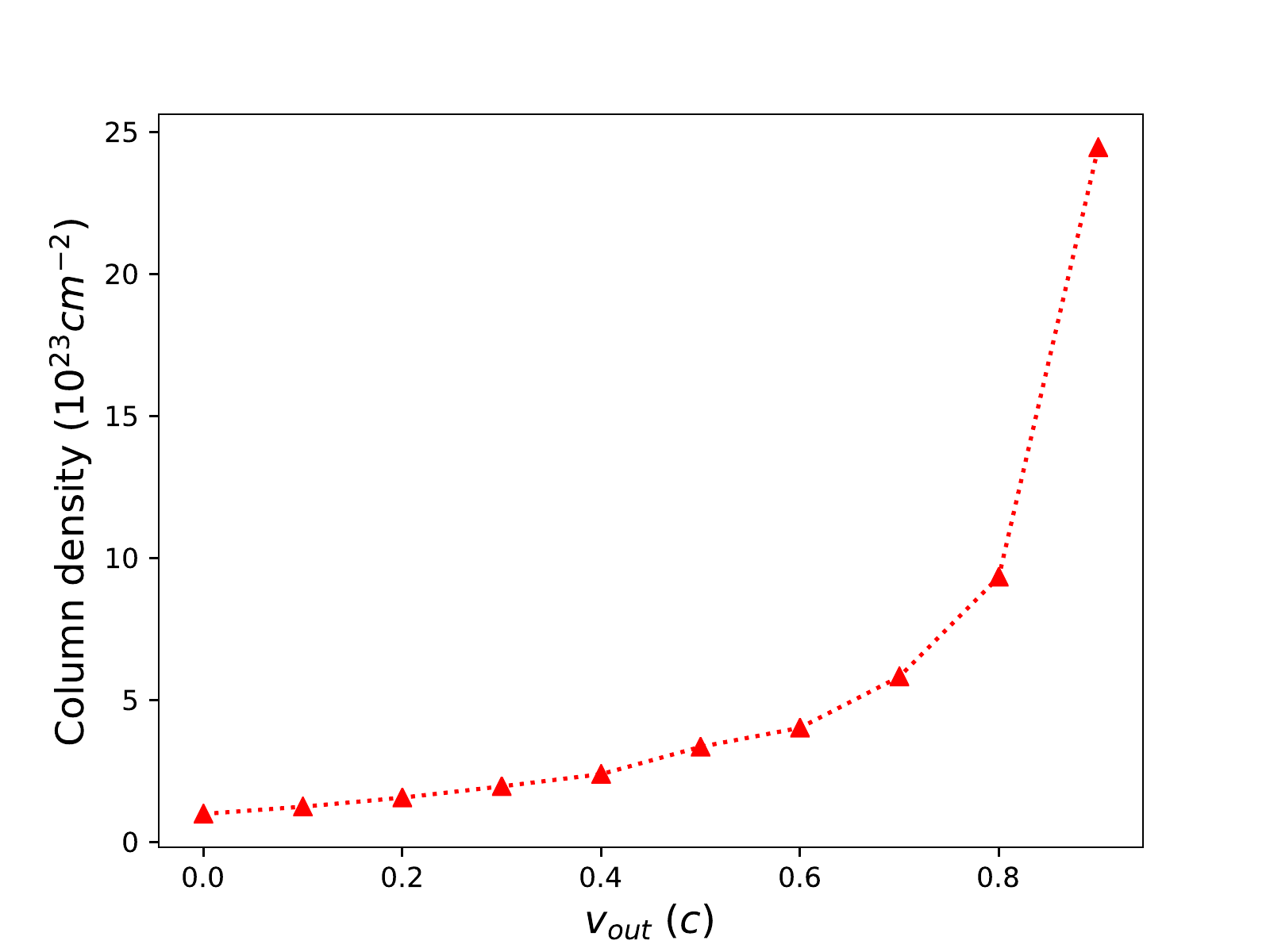}
\caption{Absorbing gas $N_H$ required to reach a given value of the optical depth as a function of $v_{out}$. Spectral parameters are as in Fig. \ref{abs_spectra}.}
\label{integral}
\end{figure}

Figure \ref{abs_spectra} shows the X-ray spectrum in the range $6-16\ keV$ of a power-law continuum source with $\Gamma=2$ and a ionising luminosity $L_{ion}=5\cdot 10^{46}\ erg\ s^{-1}$ in the 1-1000 Ry (1 Ry$= 13.6\ eV$) energy interval, modified by an absorber with $v_{out} = 0.0, 0.3$ and $0.5\ c$. In all cases, we assume an absorbing column density of $N_H=6 \cdot 10^{23} cm^{-2}$ and ionisation parameter $log(\frac{\xi}{erg\ cm\ s^{-1}})=4.5$, which are typical of Ultra Fast Outflows (UFOs) observed in AGNs (\citealp{R09, T11, G13}).
The middle and right panel of Fig. \ref{abs_spectra} also report the $v_{out}=0$ case for an easier comparison.
It can be seen that the absorption features related to the relativistically outflowing gas are both blueshifted and significantly weaker than the $v=0$ case. This effect dramatically increases for increasing velocity, as shown in a more quantitative way in Fig. \ref{integral}, which displays the column density $N_H$ necessary to reproduce outflow absorption features with a fixed optical depth, as a function of $v_{out}$. The required column density corresponds to $N_H=10^{23} cm^{-2}$ for $v_{out}=0$,  and it increases by an order of magnitude for $v_{out}=0.8c$.
It is interesting to note that this effect may introduce an observational bias in current X-ray data, which are typically restricted to $E<10 keV$, making outflows at higher velocity more difficult to detect due to the weakening of their spectral features at $E<10 keV$.

We also note that \citet{S07} and \citet{S11} presented AGN outflow models including special relativistic effects to provide an estimate of both $N_H$ and $\xi$. However, both studies seem not to account properly for the reduction of the optical depth in the calculation of $S_{out}(K)$.
Indeed, in Eq. \ref{sout} the relativistic-corrected optical depth of the wind is preserved by transforming the transmitted spectrum back to the source rest frame $K$, while this aspect has not been considered in these studies.

\section{Discussion}
\label{conclusions}

Mass and kinetic energy transfer rates of the outflow (i.e., $\dot{M}_{out}, \dot{E}_{out}$, respectively), linearly depends on $N_H$. Specifically, $\dot{M}_{out}$ can be calculated as follows (\citealp{C12}):
\begin{equation}
\dot M_{out} = 4 \pi r N_H \mu m_p C_f v_{out}
\label{Mdot}
\end{equation}
where $r, \mu, m_p, C_f$ are the distance from the source, the mean atomic weight ($\approx1.4$ for solar abundances), the proton mass and the covering factor of the outflow, respectively. $\dot{E}_{out}$ is defined as $\dot{E}_{out}=\frac{1}{2} \dot{M}_{out} v_{out}^2$. Correct estimates of $\dot{M}_{out}$ and $\dot{E}_{out}$ are of fundamental importance to test theoretical models of two-phase expansion of AGN outflows towards galaxy scales, in which kpc-scale galactic outflows are the results of the shock of ultra-fast, accretion disk-scale outflows onto the ISM (\citealp{F12, Z12, Menci19}).

\begin{figure}
\includegraphics[width=9.5cm]{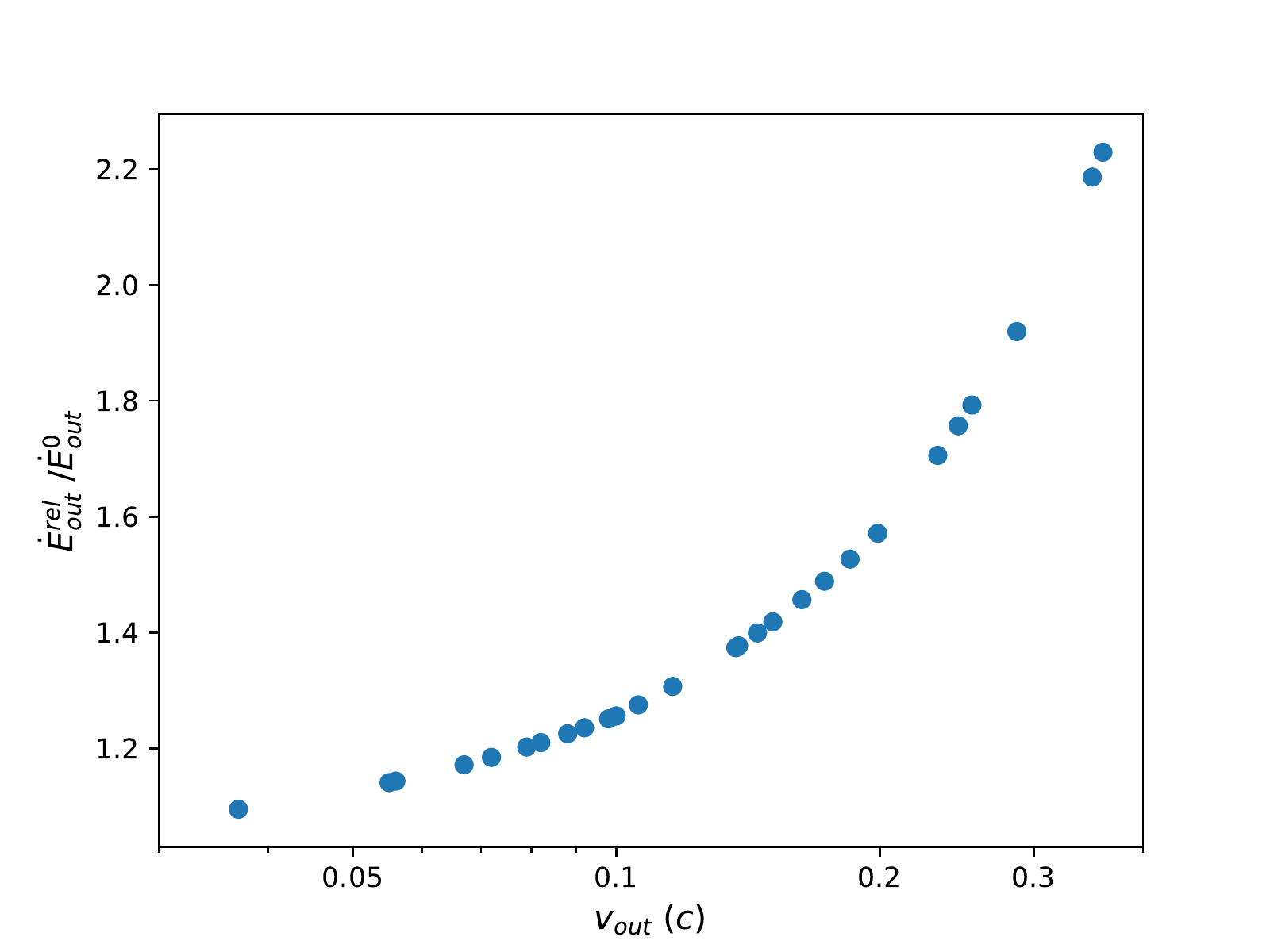}
\caption{Ratio between the relativistic-corrected energy transfer rates, $\dot{E}^{rel}_{out}$, and the original values $\dot{E}^0_{out}$ as a function of $v_{out}$, for a sample of Ultra Fast Outflows observed in AGNs (see Sect. \ref{conclusions} for details).}
\label{edot}
\end{figure}

We find that neglecting special relativistic effects will result into an underestimate of $N_H$ and, in turn, of $\dot{M}_{out},\dot{E}_{out}$. As an example, we correct for these effects the reported values of $\dot{M}_{out},\dot{E}_{out}$ for the UFOs observed in AGNs from \citet{G15} and \citet{F17} (see Figure \ref{integral}). Specifically, for the UFOs in \citet{G15} we use the average values between the reported $\dot{M}_{out},\dot{E}_{out}$ calculated using $r_{min}$ and those using $r_{max}$, where $r_{min}$ ($r_{max}$) is the minimum (maximum) inferred launching radius. Values of $\dot{M}_{out},\dot{E}_{out}$ reported in \citet{F17} are calculated in the same way. In Figure \ref{edot} we plot the ratio between the relativistic-corrected energy rates, $\dot{E}_{out}^{rel}$, and the original values, $\dot{E}_{out}^0$, as a function of $v_{out}$. $\dot{E}_{out}^{rel}$ is a factor of $>2$ higher than $\dot{E}_{out}^0$ for the fastest observed outflows ($v_{out}\geq 0.3 c$).

As shown in Fig. \ref{integral}, we expect even higher ratios for higher velocity outflows. In this respect, the improved sensitivity and resolution of the new generation X-ray telescopes, such as {\it XRISM} and {\it Athena}, will be particularly promising and it will allow to partially alleviate the observational bias discussed in Sect. \ref{prescription}. Interestingly, evidences for velocities $\geq 0.4-0.5c$ have indeed already been reported for some high luminosity quasars, such as PDS 456 and APM 08279+5255 (see e.g. \citealp{PDS, APM}).

\begin{figure}
\includegraphics[width=9.5cm]{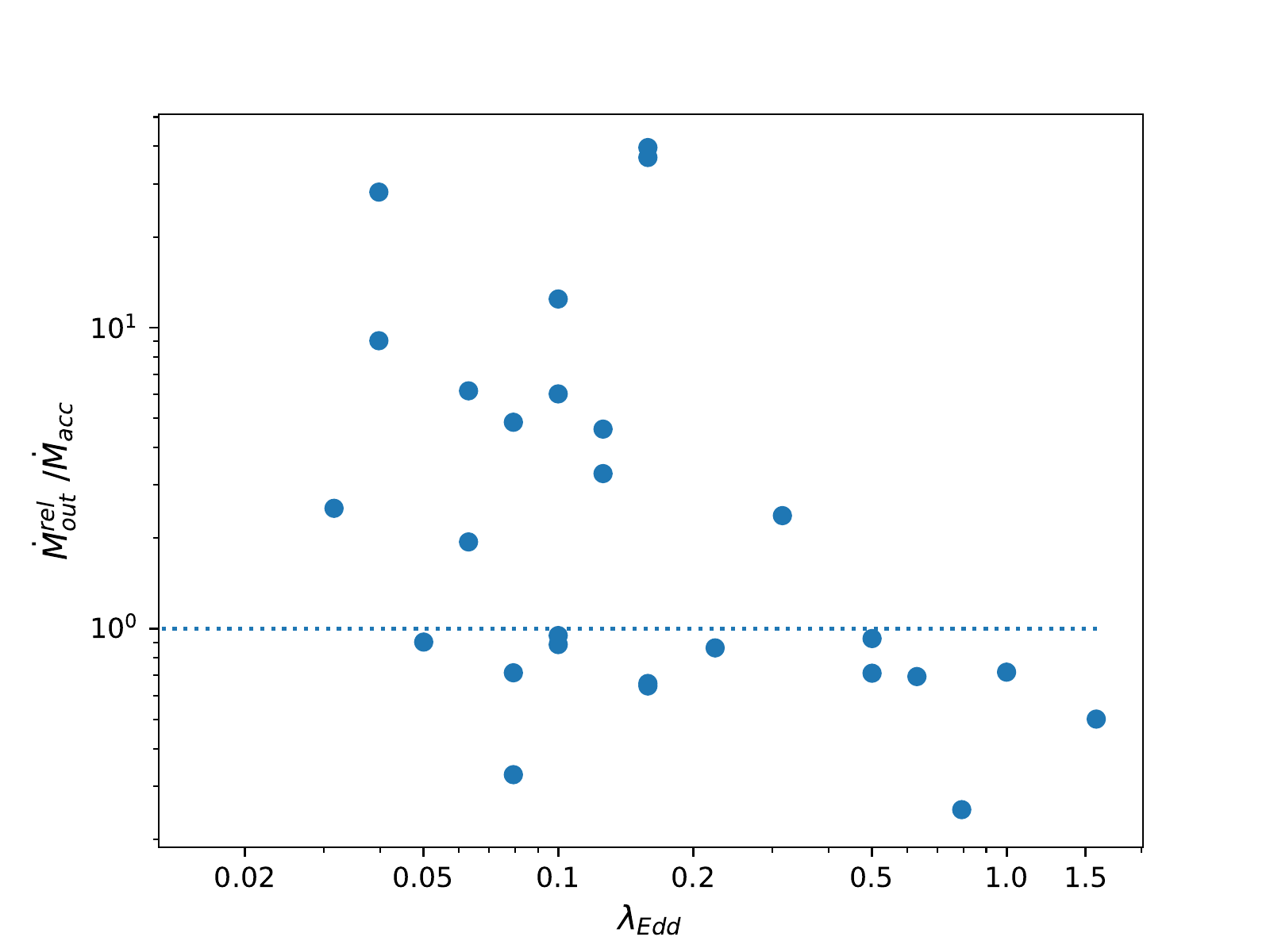}
\caption{Ratio between the relativistic-corrected outflow mass rate $\dot{M}_{out}^{rel}$ and the inflow mass rate $\dot{M}_{acc}$, as a function of $\lambda_{Edd}\equiv L_{bol}/L_{Edd}$. The sample is as in Fig. \ref{edot}. Dotted line corresponds to $\frac{\dot{M}_{out}^{rel}}{\dot{M}_{acc}}=1$.}
\label{ratio}
\end{figure}

Figure \ref{ratio} shows the ratio between the relativistic-corrected mass loss rate, $\dot{M}_{out}^{rel}$, and the mass accretion rate $\dot{M}_{acc}$, as a function of $\lambda_{Edd}\equiv L_{bol}/L_{Edd}$, i.e., the ratio between bolometric and Eddington luminosities. We derive the mass accretion rate as $\dot{M}_{acc}=\frac{L_{Bol}}{\eta c^2}$, assuming $\eta=0.1$ as in a standard \citet{SS73} accretion disk. We note that for almost half of the sources $\frac{\dot{M}_{out}^{rel}}{\dot{M}_{acc}}\geq1$, indicating that $\dot{M}_{out}^{rel}$ is comparable to (or higher than) the mass accretion rate of the disk. This may indicate a limit for the outflow lifetime, after which the accretion disk is depleted and it can no longer sustain the outflow (see e.g. \citealp{B97}). The plot also shows an apparent lack of sources with $\frac{\dot{M}_{out}^{rel}}{\dot{M}_{acc}}>1$ at large $\lambda_{Edd}$. However, the sample is too small to allow us to draw any conclusion. Future observations of high $\lambda_{Edd}$ AGNs are needed to shed light on this aspect.

Finally, we compare the outflow momentum rate, defined as $\dot{P}_{out} = \dot{M}_{out} v_{out}$, with the momentum rate of the radiation of the AGN, i.e. $\dot{P}_{rad}= \frac{L_{bol}}{c}$. We obtain a median $\frac{\dot{P}_{out}}{\dot{P}_{rad}}$ value of 0.64 for the original sample, and 0.96 after the relativistic correction. Interestingly, the latter value is consistent with unity, as expected for outflows accelerated through the continuum radiation pressure (the so-called "Eddington winds", see e.g. \citealp{KP15}).

It is worth noting that in Eq. \ref{main_long} we use the total gas velocity $v_{out}$. We assume that the velocity $v_{los}$ along the line-of-sight (LOS) coincides with $v_{out}$. As a result, the derived relativistic correction must be regarded as a conservative limit. In fact, the correction would increase in presence of an additional velocity component $v_{\perp}$ perpendicular to the LOS, which implies $v_{out}=\sqrt{v_{los}^2+v_{\perp}^2}$.

As an example, we consider the MHD model presented by \citet{Fuku10} and \citet{Fuku14}, where the outflowing gas is launched from the accretion disk at Keplerian velocity. Close to the launching radius, most of the velocity is in the direction of the disk rotation $\phi$, and it is converted in radial velocity at higher distances (i.e., close to the Alfven point) thanks to MHD effects. For a wind launched at $r_0= 10 r_G$\footnote{The gravitational radius $r_G$ is defined as $r_G=GM/c^2$, where $G$ is the gravitational constant and $M$ is the black hole mass.}, the rotational speed has a roughly constant value of $v_{\phi}=0.3 c$ until $r\approx 100 r_G$, while the radial velocity (i.e., the component parallel to the LOS) has an average value of $v_{LOS}\approx 0.2c$. 
In Figure \ref{psi} we show that when $v_{out}= 0, \Psi=1$ and the relativistic effects are absent. On the other hand, when $v_{out}\rightarrow c, \Psi \approx 0$ and the effects are the highest.
Using $v_{LOS}$ as a proxy for $v_{out}$ in Eq. \ref{main_long} yields $\Psi=0.8$, while using the total velocity (i.e., the composition between $v_r$ and $v_{\phi}$) gives $\Psi=0.6$, a factor 0.25 lower.

Noteworthy, \citet{A91} already pointed out that the observed optical depth of the gas depends on the velocity of the outflow relative to the source of radiation (see their Eqs. 2.1, 2.2). Specifically, they concentrated on an outflowing wind which is optically thick with respect to Thompson scattering, and calculated the integrated luminosity of its photosphere. Moreover, \citet{Sumi07}  and \citet{S09} considered the impact of special relativistic effects on the emitted radiation from a fast, spherical wind in stars and accreting sources, such as quasars and ULXs.
These works further underline the importance of relativistic effects for radiation-matter interaction at high speeds, along with the photo-electric and resonant absorption we investigate in this work.

\section{Conclusions}
\label{sect5}
In this work we show that special relativistic effects are of fundamental importance for a correct modellisation of the outflow spectral features, even for mildly relativistic velocities ($v_{out} \buildrel > \over \sim 0.05 c$, see Figure \ref{psi}). We also provide a simple procedure, that can be implemented in any radiative transfer code, to take into account these effects.

We observe a significant reduction of the optical depth of the outflowing gas for fixed $N_H$ and increasing $v_{out}$ (Figs. \ref{abs_spectra} and \ref{integral}). This indicates that it is necessary to include a velocity-dependent correction when estimating $N_H$ of the outflow from the optical depth derived by spectral fitting. Such correction is already significant (a factor $\approx 0.5$) for an outflow velocity of $v_{out}=0.1 c$ and reaches values a factor of $\buildrel > \over \sim 10$ for $v_{out} \geq 0.8 c$ (see Fig. \ref{integral}). 

The derived mass and kinetic energy transfer rates linearly depend on $N_H$ (see Eq. \ref{Mdot}), and hence must be corrected accordingly. For AGN outflows, this correction can significantly increase both $\dot{M}_{out},\dot{E}_{out}$ and, in turn, the impact of the outflow onto the surrounding environment, and on the feedback mechanism. We plot in Figure \ref{edot} and \ref{ratio} the relativistic-corrected $\dot{E}_{out},\dot{M}_{out}$ for a sample of Ultra Fast Outflows in AGNs reported in the literature. These pictures further underline the importance of relativistic corrections for a correct assessment of the outflow properties. Furthermore, these corrections are increasingly important in view of the next generation, high-sensitivity X-ray telescopes, which will increase the accuracy of the detection of mildly relativistic outflows, as discussed in Sect. \ref{conclusions}.

The effects discussed in Sect. \ref{physics} have further implications on the radiative driving exerted on the outflowing gas, which will be discussed in a separate work (Luminari et al., \textit{in prep}).
Moreover, we also plan to present a new version of the X-ray spectral modelling code WINE \citep{AL}, which includes a relativistic-corrected radiative transfer treatment according to the procedure of Sect. \ref{prescription}.

\emph{Acknowledgements.} We thank Stefano Ascenzi for useful discussions and Tim Kallman for having provided custom \textit{XSTAR} packages. AL, EP, FT, LZ  acknowledge financial support from the Italian Space Agency (ASI) under the contract ASI-INAF  n.2017-14-H.0. FT acknowledges support by the Programma per Giovani Ricercatori - anno 2014 “Rita Levi Montalcini”. FF acknowledges support from INAF under the contract PRIN-INAF-2016 FORECAST, and ASI/INAF contract I/037/12/0.

\begin{appendix}
\section{Algorithm for special relativity corrections}
\label{appendix1}

In this Appendix we provide a detailed description of the procedure for special relativity corrections outlined in Sect. \ref{prescription}. 

Following Eq. \ref{expl}, the incident spectrum in the outflowing gas reference frame, $S_I(K')$, is obtained from $S_I(K)$ by multiplying the frequencies by a factor $\psi$ and the intensity by a factor $\psi^3$.

For a given set of outflow parameters ($N_H, log(\xi_0), n_0, v_{out}$), we run radiative transfer simulation by using $S_I(K')$ as incident spectrum. As a result we obtain the transmitted spectrum, $S_T(K')$, displaying the absorption features due to the outflowing gas\footnote{Moreover, in some of the most popular codes, such as \textit{XSTAR}, the emissivity of all the atomic lines listed in the atomic database is saved in a separated file. Emissivities can be used to build relativistic-corrected outflow emission spectra, as we will illustrate in detail in a forthcoming paper.}.

We then calculate the "difference spectrum" as follows:
\begin{equation}
S_{T-I}(K')=S_T(K')-S_I(K')
\label{diff_spectrum}
\end{equation}
Accordingly, $S_{T-I}(K')$ represents the absorption features produced by the outflowing gas, with the relativistic-corrected optical depth. As a next step, we calculate the relativistic-corrected, rest-frame absorbed spectrum as follows:
\begin{equation}
S_{out}(K)=S_I(K)+S_{T-I}(K')\cdot \psi^{-1}
\label{eq_step}
\end{equation}
where $S_{T-I}(K')\cdot \psi^{-1}$ represents the "difference spectrum" in rest frame ($K$) frequencies, which is obtained by dividing the frequencies by a factor $\psi$. 
Using Eq. \ref{diff_spectrum}, we can thus rewrite Eq. \ref{eq_step} as:
\begin{equation}
S_{out}(K)= S_I(K)\cdot \Delta+S_T(K') \cdot \psi^{-1}
\end{equation}
where $\Delta\equiv 1-\psi^3$ and $S_I(K)\cdot \Delta$ indicates a scaling of the intensity of the spectrum $S_I(K)$ of a factor $\Delta$.

In our calculations we assume that the outflow has a net velocity $v_{out}$ and direction $\theta$. From a physical point of view, $v_{out}$ and $\theta$ correspond to the average velocity and direction of the outflow, respectively. Therefore, if a turbulent velocity component is present, the above discussion is still valid, provided that $v_{turb}\ll v_{out}$. Furthermore, if the outflowing velocity is a function of the spatial coordinates, i.e. $v_{out}=v(\overrightarrow{r})$, the above procedure can be implemented by dividing the outflow in small slabs, and assuming $v_{out}$ to be constant in each of them. Finally, the treatment of more complicated scenarios for $v(\overrightarrow{r})$ requiring a first-principle approach are beyond the scope of the present paper.
  
\end{appendix}

\end{document}